\documentclass[fleqn,usenatbib]{mnras}

\usepackage{newtxtext,newtxmath}

\usepackage[T1]{fontenc}

\DeclareRobustCommand{\VAN}[3]{#2}
\let\VANthebibliography\thebibliography
\def\thebibliography{\DeclareRobustCommand{\VAN}[3]{##3}\VANthebibliography}

\usepackage{graphicx}	
\usepackage{amsmath}	

\title[Predicting asteroid-planet close approaches]{Prediction for close approaches with terrestrial planets of asteroids from the main belt}

\author[Y. F. Zhou et al.]{
Yufan Fane Zhou,$^{1,2}$\thanks{E-mail: yufanz@smail.nju.edu.cn (YFZ)}
Zhiyuan Li,$^{1,2,3}$\thanks{E-mail: lizy@nju.edu.cn (ZL)}
Hailiang Li$^{4}$
and Liyong Zhou$^{1,2}$
\\
$^{1}$School of Astronomy and Space Science, Nanjing University, Nanjing 210046, China\\
$^{2}$Key Laboratory of Modern Astronomy and Astrophysics (Nanjing University), Ministry of Education, Nanjing 210046, China\\
$^{3}$Institute of Science and Technology for Deep Space Exploration, Suzhou Campus, Nanjing University, Suzhou 215163, China\\
$^{4}$State Key Laboratory of Lunar and Planetary Sciences, Macau University of Science and Technology, Macau 999078, China
}

\date{Accepted 2025 February 18. Received 2025 February 18; in original form 2025 January 09}

\pubyear{2025}

\begin{document}
\label{firstpage}
\pagerange{\pageref{firstpage}--\pageref{lastpage}}
\maketitle

\begin{abstract}
Potentially Hazardous Asteroids (PHAs), a special subset of Near-Earth Objects, are both dangerous and scientifically valuable. PHAs that truly undergo close approaches with the Earth (dubbed CAPHAs) are of particular interest and extensively studied. The concept and study of CAPHA can be extended to other Solar System planets, which have significant implications for future planet-based observations and explorations. In this work, we conduct numerical simulations that incorporate the Yarkovsky effect to study the transformation of main belt asteroids into CAPHAs of terrestrial planets, using precise nominal timesteps, especially to ensure the reliability of the results for Mercury and Venus. Our simulations predict a total of 1893 Mercury-CAPHAs, 3014 Venus-CAPHAs, 3791 Earth-CAPHAs and 18066 Mars-CAPHAs, with an occurrence frequency of about 1, 9, 15 and 66 per year, respectively. The values for Mars-CAPHAs are consistent with our previous work, which were based on simulations with a larger nominal timestep. The predicted occurrence frequency and velocity distribution of Earth-CAPHAs are in reasonable agreement with the observed population of Earth-CAPHAs. We also find that certain asteroids can be caught in close approach with different planets at different times, raising an interesting possibility of using them as transportation between terrestrial planets in the future.
\end{abstract}



\section{Introduction}\label{sec:introduction}
Near-Earth objects (NEOs) are small bodies with perihelion distances $q\leq 1.3$\,au and orbits approach or cross that of Earth, including both asteroids (NEAs) and comet nuclei (NECs). \citet{marsden1997} found that the orbits of NEAs can change by up to 0.05\,au over a century, and larger NEAs pose a greater threat to Earth. Therefore, NEAs with a minimum orbit intersection distance (MOID) to Earth orbit of less than 0.05\,au and an absolute magnitude $H\leq 22$\,mag (i.e., diameter $D\geq 140$\,m) are defined as potentially hazardous asteroids (PHAs) \citep{perna2013}. However, not all PHAs are truly `hazardous', as proximity in orbit does not imply proximity in actual distance (e.g. the case of Earth Trojan asteroids). Therefore, PHAs that actually undergo close approaches (hereafter called CAPHAs) to the Earth are the key targets of human monitoring\footnote{https://cneos.jpl.nasa.gov/ca}, in which case the asteroids have a minimum distance $r_{\rm min} <$ 0.05\,au from Earth.

CAPHAs (or more generally, NEAs) are worth studying not only for safety reasons but also for scientific purposes. On one hand, potential impact events can pose significant threats to the Earth, such as the `K-T' event leading to the Cretaceous/Tertiary extinction that occurred about 65 million years ago \citep{kyte1998,bottke2007}, the Tungus Explosion in 1908 \citep{ben1975}, and the Chelyabinsk meteor falling in 2013 \citep{brown2013}. On the other hand, NEAs could have delivered water and organic-rich materials to the early Earth, which is crucial for our understanding about the origin of life as well as the early evolution of the Solar System \citep{marty2012,alexander2012,altwegg2015}.

The primary sources of NEAs are unstable regions in the main asteroid belt \citep{wetherill1988,rabinowitz1997a,rabinowitz1997b,bottke2002}, such as the $\nu_6$ secular resonance \citep{morbidelli1994} and the 3:1 mean motion resonance with Jupiter \citep{wisdom1983,moons1996}, which manifest as `gaps' in the orbital distribution of main belt asteroids. In addition to gravitational forces, the Yarkovsky effect, which can push nearby asteroids into gaps, also play a significant role in transporting asteroids into the inner Solar System \citep{burns1979,rubincam1995,farinella1998,bottke2006}. Due to an asteroid's rotation, revolution around the Sun, and finite thermal inertia, it has an asymmetric surface temperature distribution. The Yarkovsky effect indicates that the net recoil force generated by an asteroid's asymmetric thermal re-radiation can alter its orbit over a reasonable long time scale \citep{peterson1976,farinella1998,farinella1999}. Such radiation force can also produce a torque that modifies the spin rate and axis orientation of the asteroid, which is referred to as the Yarkovsky-O'Keefe-Radzievskii-Paddack (YORP) effect \citep{bottke2006}.

The above concepts and theories can be applied not only to asteroids near Earth but also to those near other terrestrial planets. Similar to Earth-CAPHAs, the definition of CAPHAs for Mercury (Venus) (Mars) is asteroids with an absolute magnitude $H\leq 22$\,mag and a minimum distance $r_{\rm min} <$ 0.007 (0.034) (0.036)\,au from the planet, as the Hill radius of Mercury (Venus) (Mars) is 0.15 (0.67) (0.72) times that of Earth. Although the distance criterion in the original definition of Earth-CAPHAs was not derived from the Hill radius \citep{marsden1997}, it is reasonable to set different values for different planets, taking into account their varying abilities to capture nearby small bodies. With the advancement of aerospace technology, more and more planetary exploration missions are underway or upcoming, such as BepiColombo \citep{benkhoff2010}, DAVINCI \citep{garvin2022}, Mars~2020 \citep{farley2020} and Tianwen-1 \citep{li2021}. These missions make it possible to image and analyze CAPHAs of Mercury, Venus and Mars, which will contribute to our understanding of planetary environments and their evolutionary histories.

Therefore, numerical simulations about CAPHAs that can provide guidance for observations are timely and necessary. Earth-CAPHAs and Mars-CAPHAs have been investigated by simulations in \citet{zhou2024}. Here, following a similar approach, we conduct \textit{N}-body simulations incorporating the Yarkovsky effect for the Solar System in order to study Mercury-CAPHAs and Venus-CAPHAs, while also updating the results for Earth and Mars.  The methods are presented in Section~\ref{sec:methods}, including the software, the implementation of the Yarkovsky effect, the data used in this work, and the simulation configurations. In Section~\ref{sec:results}, we show the results for CAPHAs of different planets and make some discussions. Section~\ref{sec:summary} presents a brief summary.

\section{Methods}\label{sec:methods}
\subsection{Simulation tool with the Yarkovsky force subroutine}
We use \textsc{Mercury6} \citep{chambers1999}, a general-purpose software package for doing \textit{N}-body integrations, as our simulation tool. The integrator we chose is the hybrid integrator, which includes both the symplectic algorithm and the Bulirsch-Stoer algorithm. The Bulirsch-Stoer integrator (an integrator with adaptive timesteps) will only take effect when a planetary close encounter occurs. We incorporate the Yarkovsky effect into the subroutine \textit{mfo\_user} using the same approach as \citet{zhou2024}.

The Yarkovsky effect results in recoil forces on asteroids, but the precise calculation is highly complex and dependent on a lot of physical parameters, some of which have not yet been accurately measured for many asteroids. Therefore, the Yarkovsky effect is simply described as an equivalent drift rate of the semi-major axis ${\rm d}a/{\rm d}t$ of an asteroid in some studies \citep{vokrouhlicky1998,vokrouhlicky1999}. For example, an asteroid located at $a=2.9$\,au with diameter $D=1$\,km has a drift rate of ${\rm d}a/{\rm d}t = 0.256$\,au\,Gyr$^{-1}$ \citep{xu2020}, based on the assumptions of a rotation period $P=8$\,h, albedo $p=0.13$, thermal conductivity $K=0.005$\,W\,m$^{-1}$\,K$^{-1}$, specific heat capacity $C=680$\,J\,kg$^{-1}$\,K$^{-1}$, surface density $\rho_{\rm s}=1.5$\,g\,cm$^{-3}$, bulk density $\rho=2.5$\,g\,cm$^{-3}$ and spinning obliquity $\gamma=0$\,deg \citep{vokrouhlicky2006}. Although the relationship between drift rate and semi-major axis may be complicated \citep{xu2020b}, ${\rm d}a/{\rm d}t$ can be considered inversely proportional to $a^2$ within the range of the main belt \citep{nesvorny2023}. Additionally, the drift rate should be inversely proportional to the diameter $D$ \citep{xu2020b,nesvorny2023}. Based on discussions above and consistent with \citet{zhou2024}, the formula for drift rate used in our work is
\begin{equation}
    \frac{{\rm d}a}{{\rm d}t}=0.25\,\frac{\rm au}{\rm Gyr}\times \left(\frac{1\,{\rm km}}{D}\right)\left(\frac{2.958\,{\rm au}}{a}\right)^{2}.
	\label{eq:dadt}
\end{equation}

\subsection{Sampling and grouping of main belt asteroids}
\begin{figure}
    \centering
    \includegraphics[width=\columnwidth]{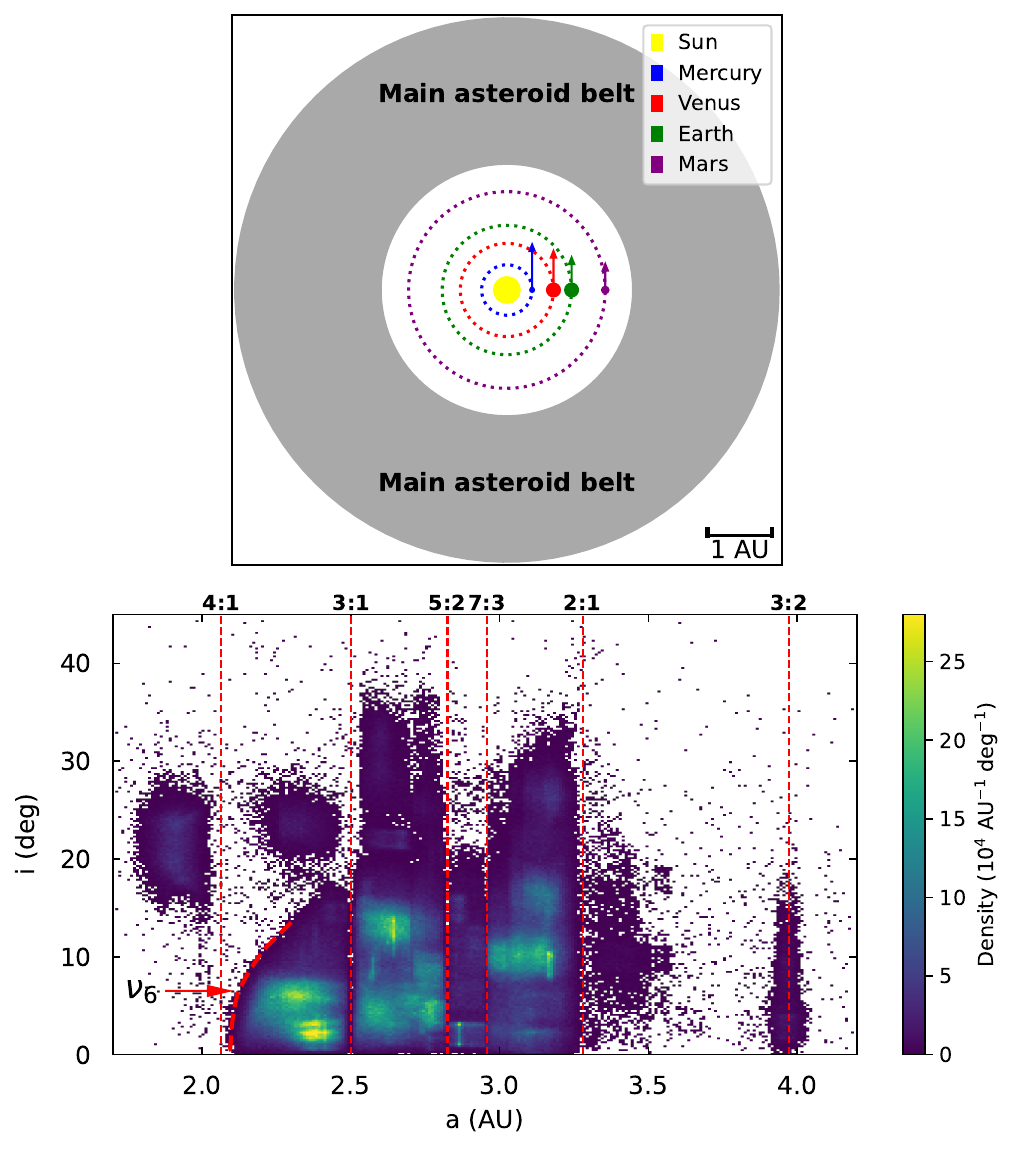}
    \caption{Upper panel: A schematic representation of the Sun, four terrestrial planets, and the main asteroid belt. Lower panel: Scatter plot of inclination $i$ versus semi-major axis $a$ of the main belt, plotted using data from the Small-Body Database of Jet Propulsion Laboratory. The $\nu_6$ secular resonance and several mean motion resonances are marked with red dashed lines.}
    \label{fig:1}
\end{figure}

CAPHAs of Mercury, Venus, Earth and Mars should mainly originate from the main asteroid belt \citep{rabinowitz1997a,rabinowitz1997b}, which has been stable for a considerable period of time. During this period, the escape of asteroids does not significantly change the distribution of the main belt. Based on such assumptions, we use actually observed data of main belt asteroids as the initial conditions of small bodies for our simulations. Data for 1,223,263 main belt asteroids (as of June 26, 2023) are downloaded from the Small-Body Database\footnote{https://ssd.jpl.nasa.gov/tools/sbdb\_query.html} (SBDB) of Jet Propulsion Laboratory (JPL). Their distribution is shown in the lower panel of Figure~\ref{fig:1}. After removing those with large errors and randomly sampling 1\%, our `sample data' contains 11445 asteroids, which is the same as that used in \citet{zhou2024}.

The 11445 asteroids are divided into two categories named `near-gap' asteroids (709) and `far-gap' asteroids (10736). Specific ranges of `near-gap' regions are shown in column~3 of Table~\ref{tab:setting_result}, and the methods for determining them are as follows. Given that the width of the 7:3 resonance (2.958\,au) is explicitly identified to be 0.01\,au \citep[e.g.][]{vokrouhlicky2006,xu2020}, we set the width of this `near-gap' region to 0.015\,au, 1.5 times the resonance width, to ensure that its range is significantly larger than the resonance size. For the other five mean motion resonances considered by us, we set the width of their corresponding `near-gap' regions such that the number of asteroids they encompass is comparable to that of the 7:3 resonance. The widths determined here follow the trend that lower-order resonances have wider regions, which reflects the reasonableness of our approach. It should be noted that the 4:1 resonance is quite close to the $\nu_6$ resonance, so we set a relatively large width for it and extend its right edge to 2.200\,au \citep{wetherill1985} to embrace the $\nu_6$ resonance \citep{scholl1991}. In addition to the six mean motion resonances shown in the lower panel of Fig.~\ref{fig:1}, there are some other relatively strong ones, such as the 5:1 resonance (1.780\,au). However, in our sample, their regions do not contain over $\sim$10 asteroids, even when a particularly large width is set. Therefore, when studying `near-gap' asteroids, we only consider the six regions mentioned before.

\begin{table*}
\footnotesize
	\centering
	\caption{The basic information of each simulation and the results for Mercury-CAPHAs and Venus-CAPHAs. Columns 4 and 5 show the number of CAPHAs directly output by the simulations, while columns 9 and 10 display the numbers corrected using the method described in Section~\ref{sec:results_quantity}. The numbers 1, 2, 3, 4, 5, 6, and 7 in the ID of a simulation correspond to $H$ (mag) of 15.5, 16.5, 17.5, 18.5, 19.5, 20.5, and 21.5, or equivalently $D$ (km) of 2.82, 1.78, 1.12, 0.71, 0.45, 0.28, and 0.18, respectively, in the simulation. `MC' and `VC' are abbreviations for Mercury-CAPHA and Venus-CAPHA. The range (au) of series F (3:2 resonance) is $3.972\pm 0.050$, whose information is not shown as it contributes insignificantly to the results.}
	\label{tab:setting_result}
	\begin{tabular}{cccccccccc}
		\hline
		ID & drift rate (au\,Gyr$^{-1}$) & region information & $N_{\rm MC}$ & $N_{\rm VC}$ & $\bar \epsilon_{\rm MC}$ & $\bar \epsilon_{\rm VC}$ & $\eta$ & $N_{\rm MC,obs}$ & $N_{\rm VC,obs}$ \\
            (1) & (2) & (3) & (4) & (5) & (6) & (7) & (8) & (9) & (10) \\
		\hline
        A1 & 0.18  & range (au): & 41 & 44 & 0.043 & 0.073 & 0.051 & 0.089 & 0.164\\
        A2 & 0.29 & $2.065^{+0.135}_{-0.050}$ & 51 & 59 & 0.029 & 0.049 & 0.115 & 0.167 & 0.334\\
        A3 & 0.46 & ~ & 47 & 48 & 0.053 & 0.080 & 0.261 & 0.650 & 1.000\\
        A4 & 0.72 & resonance: & 65 & 71 & 0.044 & 0.056 & 0.236 & 0.668 & 0.933\\
        A5 & 1.13 & 4:1 and $\nu_6$ & 53 & 59 & 0.103 & 0.110 & 0.355 & 1.940 & 2.307\\
        A6 & 1.82 & ~ & 72 & 77 & 0.023 & 0.037 & 0.535 & 0.881 & 1.540\\
        A7 & 2.83 & population: 198 & 70 & 78 & 0.045 & 0.062 & 0.806 & 2.532 & 3.896\\
        \hline
        B1 & 0.12 & range (au): & 38 & 44 & 0.031 & 0.031 & 0.079 & 0.093 & 0.109\\
        B2 & 0.20 & $2.502\pm 0.025$ & 52 & 58 & 0.021 & 0.024 & 0.198 & 0.221 & 0.272\\
        B3 & 0.31 & ~ & 61 & 74 & 0.008 & 0.010 & 0.497 & 0.252 & 0.378\\
        B4 & 0.49 & resonance: & 58 & 71 & 0.012 & 0.015 & 0.483 & 0.332 & 0.509\\
        B5 & 0.78 & 3:1 & 63 & 82 & 0.010 & 0.011 & 0.765 & 0.484 & 0.701\\
        B6 & 1.25 & ~ & 79 & 91 & 0.023 & 0.031 & 1.212 & 2.187 & 3.437\\
        B7 & 1.94 & population: 122 & 86 & 97 & 0.007 & 0.010 & 1.919 & 1.117 & 1.853\\
        \hline
        C1 & 0.10 & range (au): & 10 & 28 & 0.000 & 0.003 & 0.120 & 0.000 & 0.009\\
        C2 & 0.15 & $2.825\pm 0.020$ & 7 & 31 & 0.003 & 0.002 & 0.322 & 0.007 & 0.022\\
        C3 & 0.24 & ~ & 11 & 33 & 0.016 & 0.007 & 0.868 & 0.151 & 0.190\\
        C4 & 0.38 & resonance: & 8 & 45 & 0.038 & 0.008 & 0.885 & 0.269 & 0.334\\
        C5 & 0.60 & 5:2 & 12 & 42 & 0.004 & 0.003 & 1.453 & 0.068 & 0.181\\
        C6 & 0.96 & ~ & 20 & 69 & 0.002 & 0.003 & 2.384 & 0.118 & 0.437\\
        C7 & 1.50 & population: 132 & 18 & 72 & 0.005 & 0.004 & 3.913 & 0.383 & 0.997\\
        \hline
        D1 & 0.09 & range (au): & 1 & 3 & 0.000 & 0.001 & 0.120 & 0.000 & 0.000\\
        D2 & 0.14 & $2.958\pm 0.015$ & 4 & 10 & 0.001 & 0.003 & 0.322 & 0.001 & 0.009\\
        D3 & 0.22 & ~ & 1 & 6 & 0.035 & 0.009 & 0.868 & 0.031 & 0.045\\
        D4 & 0.35 & resonance: & 3 & 19 & 0.000 & 0.001 & 0.885 & 0.000 & 0.020\\
        D5 & 0.56 & 7:3 & 1 & 15 & 0.003 & 0.000 & 1.453 & 0.005 & 0.010\\
        D6 & 0.89 & ~ & 9 & 21 & 0.005 & 0.005 & 2.384 & 0.118 & 0.253\\
        D7 & 1.39 & population: 125 & 5 & 14 & 0.000 & 0.001 & 3.913 & 0.008 & 0.044\\
        \hline
        E1 & 0.07 & range (au): & 5 & 7 & 0.002 & 0.005 & 0.172 & 0.001 & 0.006\\
        E2 & 0.11 & $3.279\pm 0.050$ & 6 & 11 & 0.004 & 0.006 & 0.470 & 0.013 & 0.031\\
        E3 & 0.18 & ~ & 9 & 15 & 0.012 & 0.014 & 1.284 & 0.137 & 0.275\\
        E4 & 0.28 & resonance: & 5 & 16 & 0.007 & 0.005 & 1.322 & 0.047 & 0.111\\
        E5 & 0.44 & 2:1 & 5 & 13 & 0.017 & 0.011 & 2.185 & 0.188 & 0.306\\
        E6 & 0.71 & ~ & 10 & 22 & 0.014 & 0.009 & 3.611 & 0.499 & 0.691\\
        E7 & 1.11 & population: 77 & 11 & 20 & 0.080 & 0.073 & 5.968 & 5.276 & 8.733\\
		\hline
        G & 0 & population: 10736 & 0 & 0 & - & - & - & 0 & 0\\
		\hline
        total & - & - & - & - & - & - & - & 18.93 & 30.14\\
        \hline
	\end{tabular}
\end{table*}

\subsection{Simulation settings}
\label{subsec:setting}
Similar to previous work, the Sun, eight planets and (massless) asteroids are considered in our dynamical model. The initial conditions of eight planets at epoch JD 2457724.5 are obtained from the JPL HORIZONS system \citep{giorgini1996}. Asteroids with different epochs from planets are integrated to the epoch above automatically by \textsc{Mercury6} before the main integration begins. We set the nominal timestep to 4 hours, which has been significantly reduced compared to our previous work, allowing us to resolve even the highest-speed Mercury encounters \citep{greenstreet2012}. Readers are referred to \citet{zhou2024} for more details.

For `near-gap' asteroids, we run simulations with the Yarkovsky effect for 0.1\,Gyr, which is sufficiently long for the effect to become evident. For `far-gap' asteroids, consistent with \citet{zhou2024}, we neglect the Yarkovsky effect (i.e. only with gravity) and run simulations for 1\,Myr only, a simplification in order to save computational effort. The reasons for adopting different simulation configurations are as follows. Some work \citep[e.g.][]{granvik2017} shows that the Yarkovsky effect can deliver 100-meter-diameter asteroids from most regions in the main belt to $q=1.3$\,au within tens of millions of years, but such a delivery predicts too many NEAs compared to observations, due to the absence of the YORP effect in the simulation. However, simulations that account for both the Yarkovsky and YORP effects yield results lower than the observed values \citep{granvik2017}, possibly because the canonical YORP model still requires improvement. Therefore, there are currently difficulties in describing the drift timescale of main belt asteroids caused by the Yarkovsky and YORP effect. Nevertheless, for `near-gap' asteroids, the impact of the drift timescale is minimal, as they are initially close to strong resonances, hence considering only the Yarkovsky effect can obtain results consistent with observations \citep{xu2020b}. Moreover, since we focus on asteroids with sizes greater than 100-meter (more accurately, greater than 140-meter), which would take an even longer time to excite, `far-gap' asteroids should contribute only a minor fraction of the resultant CAPHAs. Thus, our separate treatment for `near-gap' and `far-gap' asteroids is justified.

To calculate the Yarkovsky drift rate of an asteroid, its semi-major axis $a$ and diameter $D$ are required. Here, for a `near-gap' region, $a$ of all asteroids within it are approximated by the value of the center of this region. However, $D$ of many asteroids are still unknown. Moreover, observations of the main belt are incomplete: asteroids with small $D$ (i.e., large absolute magnitude $H$) must have been missed, and such selection effect is more severe for more distant regions. Therefore, we generate multiple sets of cloned asteroids for each `near-gap' region and assign them different $H$ values in our simulations, in order to prepare for subsequent correction for the selection effect (see Section~\ref{sec:results_quantity}). $H$ and $D$ follow the empirical conversion relationship:
\begin{equation}
    D=\frac{1329\,{\rm km}}{\sqrt{p}}\times 10^{-0.2H}.
	\label{eq:hd}
\end{equation}
Here albedo $p$ is set to 0.14 \citep{mainzer2011}, which is close to the value used in the calculation of the drift rate \citep{xu2020}. Since our `far-gap' simulation does not account for the Yarkovsky effect, simulations with different $H$ values are not required, but the selection effect will still be corrected (see Section~\ref{sec:results_quantity}).

The absolute magnitude range considered by us is from 15\,mag (the number of asteroids with $H<15$\,mag is small) to 22\,mag (the upper limit of CAPHA's definition). The range is divided into 7 intervals, with each interval spanning 1\,mag. For each interval, the entire range is roughly represented by the central magnitude value (e.g., 15.5, 16.5, ...).

\begin{figure}
    \includegraphics[width=\columnwidth]{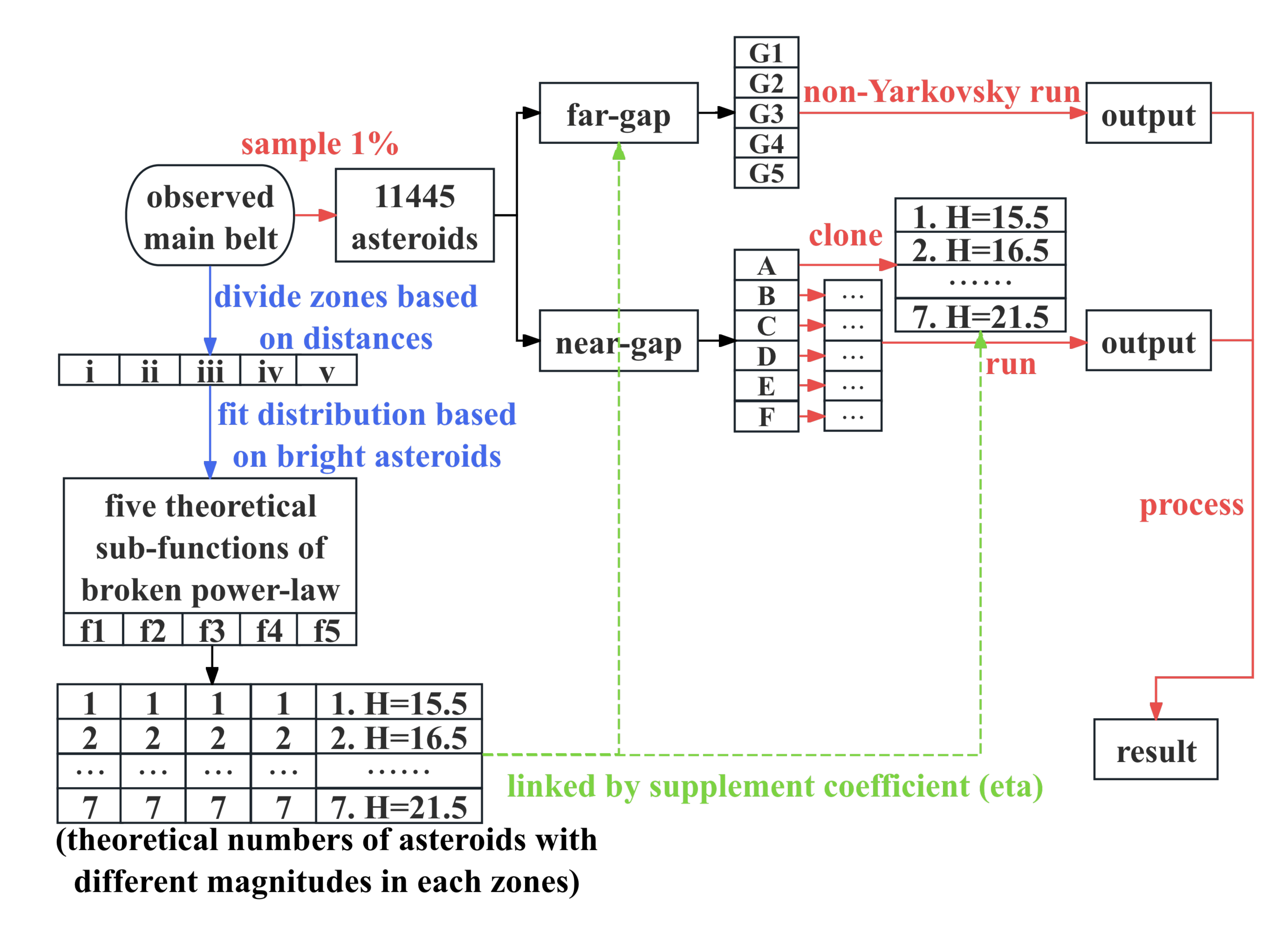}
    \caption{The flowchart of our work, which provides a simplified overview of how simulations are designed and how selection effects are corrected.}
    \label{fig:2}
\end{figure}

The specific settings of our simulations are shown in columns~1-3 of Table~\ref{tab:setting_result}, where six `near-gap' regions are designated by A-F and the `far-gap' region is designated by G. For each `near-gap' region, a total of 7 simulations are conducted, and simulations~1-7 correspond to the $H$ value ranging from 15.5 to 21.5, respectively. A simple flowchart of our work is shown in Figure~\ref{fig:2}.

It should be noted that the drift rate of a specific asteroid remains constant in our simulation, but this does not lead to distortion in the results, as the Yarkovsky effect becomes negligible compared to gravitational forces after the asteroid escapes its original region in the main belt. Furthermore, the Yarkovsky effect can either increase or decrease an asteroid's semi-major axis, depending on the direction of its rotation. However, the rotation direction of most main belt asteroids is unknown at present. Therefore, in our simulations, each asteroid is assigned a random drift direction, which is more realistic compared to the setting in \citet{zhou2024}. Although such an assignment ignores the intermediate states of asteroid obliquity, it is acceptable as a simplification.

\section{Results}\label{sec:results}
\subsection{Quantities of CAPHAs}
\label{sec:results_quantity}
The direct outputs from our simulations are shown in columns~4-5 of Table~\ref{tab:setting_result}, which are subject to further processing. First, we introduce the active coefficient
\begin{equation}
    \epsilon=\frac{t_{\rm act}}{T_{\rm sim}},
	\label{eq:epsilon}
\end{equation}
where $T_{\rm sim}$ is the simulation duration and $t_{\rm act}$ is the active lifetime. The active lifetime is the time difference between an asteroid's first and last occurrence as a CAPHA; for asteroids with just a single close approach, its active lifetime is set to be 5 years, which corresponds to the orbital period for $a=3$\,au (the center of the main belt), instead of 0. Taking a simulation spanning 100\,Myr as an example, if the active lifetime of a CAPHA is 5\,Myr, its $\epsilon$ would be 0.05. This means its contribution to the total count is not 1 but rather 0.05 (this number still requires further adjustment, see the next paragraph). Considering that listing the active coefficients for all CAPHAs generated in simulations would be too cumbersome, we only present the average active coefficient $\bar \epsilon$ for each simulation in columns~6-7 of Table~\ref{tab:setting_result}.

Second, the supplement coefficient $\eta$ is introduced to correct the observational selection effects of the main asteroid belt (see Figure~\ref{fig:2} for a quick overview). Given the different completeness at different distances, the main belt is partitioned into five zones named $\romannumeral1$, $\romannumeral2$, $\romannumeral3$, $\romannumeral4$ and $\romannumeral5$, using 2.3\,au, 2.7\,au, 3.1\,au and 3.5\,au as dividing points. Our `near-gap' simulations A, B, C\&D, E, F belong to zones $\romannumeral1$, $\romannumeral2$, $\romannumeral3$, $\romannumeral4$ and $\romannumeral5$, respectively. Simulation~G is also correspondingly subdivided into G1, G2, G3, G4, and G5 (see Table~\ref{tab:setting_result_mars}). Assuming that the actual cumulative magnitude distribution of asteroids follows a broken power-law with a break point at $H=18$ \citep{gladman2009}, we use asteroids with $15\leq H\leq H^{\prime}$ in each of the five zones to separately fit their distribution functions for $H \leq 18$ (the first part of the broken-power law). The values of $H^{\prime}$ are chosen based on the currently observed cumulative magnitude distribution and are respectively set to 18.0, 18.0, 17.3, 17.0 and 16.2. We obtain:
\begin{equation}
    lg[N(<H)]=\left\{
              \begin{aligned}
              \begin{array}{ll}
              0.36H-1.68, & \romannumeral1\,(a<2.3)\\
              0.40H-1.65, & \romannumeral2\,(2.3\leq a<2.7)\\
              0.43H-2.10, & \romannumeral3\,(2.7\leq a<3.1)\\
              0.44H-2.32, & \romannumeral4\,(3.1\leq a<3.5)\\
              0.41H-3.03, & \romannumeral5\,(a\geq 3.5)\\
              \end{array}
              \end{aligned}
              \right
              .
	\label{eq:exp}
\end{equation}
For the case of $H>18$ (the second part of the broken-power law), while ensuring the function passes through the break point, the slopes of the five sub-equations are halved \citep{gladman2009}. Then, for each zone, we can deduce the ratio of numbers for asteroids in 7 magnitude intervals (corresponding to simulations 1-7), and we define $q_{\rm xk}$ as the $k$-th value among the 7 ratios of zone $x$. For instance, the ratio for zone~$\romannumeral4$ is 1.00~:~2.73~:~7.46~:~7.68~:~12.70~:~20.98~:~34.68, where the first value ($q_{\rm \romannumeral4 1}$, set to 1.00) corresponds to 15-16\,mag, and the last value ($q_{\rm \romannumeral4 7}$) corresponds to 21-22\,mag. However, due to the difficulty in observing faint objects, the observed distributions deviate significantly from the theoretical ones mentioned above. Therefore, we introduce the supplement coefficient for zone $x$, magnitude interval $k$:
\begin{equation}
    \eta_{\rm xk}=\frac{n_{\rm x,1}}{n_{\rm x,total}}\times q_{\rm xk},\quad x=\romannumeral1-\romannumeral5,k=1-7.
	\label{eq:eta}
\end{equation}
$n_{\rm x,1}$ and $n_{\rm x,total}$ represent the number of actually observed asteroids with $15\leq H<16$ and $15\leq H<22$ in zone $x$, respectively. For a `near-gap' simulation (A--F), determining its associated zone and magnitude interval will obtain its $\eta$. For a `far-gap' simulation (G1--G5), its $\eta$ value is the average of the seven $\eta$ values (corresponding to the seven magnitude intervals) of the associated zone. Then if one simulation contributes $N$ CAPHAs, we correct its quantity to $\eta N$. The $\eta$ values of all simulations are shown in column~8 of Table~\ref{tab:setting_result} and Table~\ref{tab:setting_result_mars}.

\begin{figure}
    \includegraphics[width=\columnwidth]{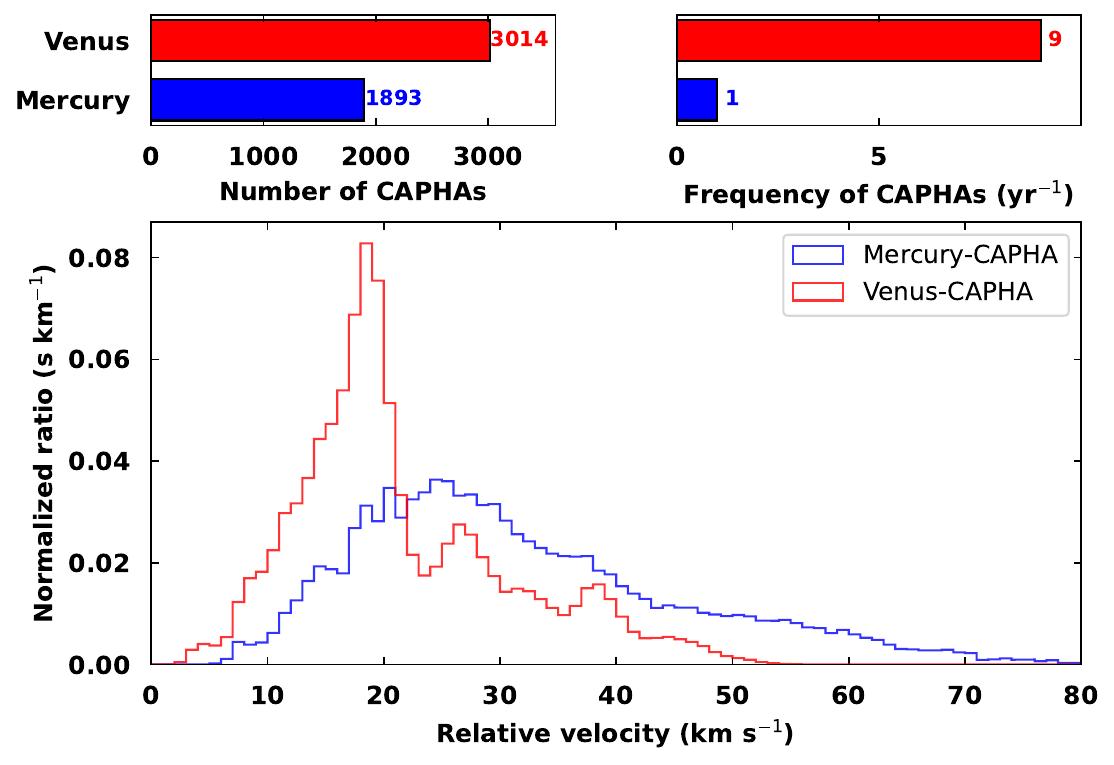}
    \caption{Upper left (right) panel: Numbers (Frequencies) of Mercury-CAPHAs and Venus-CAPHAs obtained from simulations. Lower panel: Distribution of velocities of CAPHAs relative to the planet.}
    \label{fig:3}
\end{figure}

Lastly, we multiply the quantity $N$ by coefficients $\epsilon$ and $\eta$ and sum up results of all simulations to obtain the observable number of CAPHAs:
\begin{equation}
    N_{\rm obs}=\sum\limits_{j=1}^{5}\epsilon_{\rm Gj}\eta_{\rm Gj}N_{\rm Gj}+\sum\limits_{X=A}^{F}\sum\limits_{k=1}^{7}\epsilon_{\rm Xk}\eta_{\rm Xk}N_{\rm Xk}.
	\label{eq:number}
\end{equation}
A total count of 18.93 Mercury-CAPHAs and 30.14 Venus-CAPHAs are obtained. Given that our sample represents only 1\% of the main asteroid belt, the predicted numbers of observable Mercury-CAPHAs and Venus-CAPHAs are 1893 and 3014 (see Figure~\ref{fig:3}).

Then we estimate their occurrence frequencies. We first calculate the average frequency of each CAPHA for each simulation:
\begin{equation}
    \bar f_{\rm Xk}=\frac{1}{N_{\rm Xk}}\sum\limits_{m=1}^{N_{\rm Xk}}\frac{C_{\rm Xk,m}}{t_{\rm act,Xk,m}}.
	\label{eq:frequency_f}
\end{equation}
$N_{\rm Xk}$ represents the number of CAPHAs generated in simulation~Xk (e.g., B7), $C$ denotes the count of repeated close approaches for a CAPHA, and $t_{\rm act}$ is its active lifetime. It should be noted that, as mentioned when giving equation~(\ref{eq:epsilon}), $t_{\rm act}$ of asteroids that only visit once has been set to 5 years, which can avoid the occasional occurrence of a zero denominator in equation~(\ref{eq:frequency_f}). However, we notice that not just a few, but several Mercury-CAPHAs visit Mercury only once, which could significantly skew the frequency due to the artificially assigned lifetime. Therefore, when calculating the frequency of Mercury-CAPHAs, we temporarily set their active lifetime to the simulation duration to minimize their impact on the result.

Next, for equation~(\ref{eq:number}), we multiply each term in the summation by its corresponding $\bar f$, yielding the total frequency:
\begin{equation}
    f_{\rm obs}=\sum\limits_{j=1}^{5}\epsilon_{\rm Gj}\eta_{\rm Gj}N_{\rm Gj}\bar f_{\rm Gj}+\sum\limits_{X=A}^{F}\sum\limits_{k=1}^{7}\epsilon_{\rm Xk}\eta_{\rm Xk}N_{\rm Xk}\bar f_{\rm Xk}.
	\label{eq:frequency_F}
\end{equation}
The resultant occurrence frequencies of Mercury-CAPHAs and Venus-CAPHAs are about 1 and 9 per year, respectively. Therefore, for Mercury and Venus, the chances of observing close encounter events of asteroids with sizes greater than 140-meter are relatively low.

\subsection{Updates on results for Earth and Mars}
\begin{figure}
    \includegraphics[width=\columnwidth]{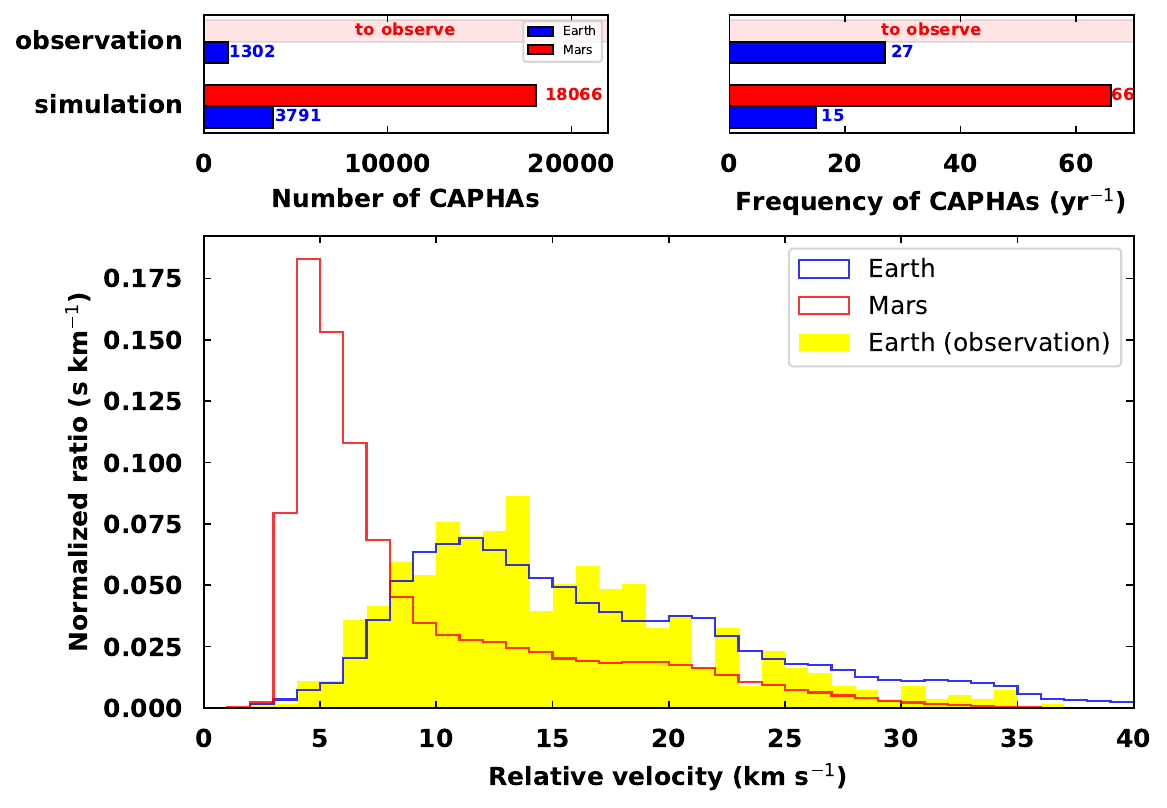}
    \caption{Upper left (right) panel: Numbers (Frequencies) of Earth-CAPHAs and Mars-CAPHAs obtained from simulations and actual observations. Lower panel: Distribution of velocities of CAPHAs relative to the planet (9 anomalous asteroids were excluded while plotting, see Section~\ref{sec:velocity} for reasons). The yellow histogram shows the actually observed distribution of Earth-CAPHAs.}
    \label{fig:4}
\end{figure}

The results for Earth-CAPHAs and Mars-CAPHAs have been presented in \citet{zhou2024}. Here, we optimize the nominal timestep and drift direction settings (see Section~\ref{subsec:setting}), and the updated results for Earth and Mars are shown in Figure~\ref{fig:4} and Table~\ref{tab:setting_result_mars}.

Our simulations predict 3791 Earth-CAPHAs and 18066 Mars-CAPHAs, i.e. a ratio of 1:4.8. As of June 26, 2023, more than 1300 independent Earth-CAPHAs (counted without repetition) have been observed and catalogued, indicating that there are still $\sim$2 times more left to be observed. The occurrence frequencies of Earth-CAPHAs and Mars-CAPHAs are 15 and 66 per year, respectively. We note that the frequency of Earth-CAPHAs above is in rough agreement with the actually observed frequency ($\sim$27 per year averaged over 20 years), in light of the potential underestimation of the contribution from `far-gap' regions (see Section~\ref{subsec:setting}). In conclusion, Mars-CAPHAs significantly surpass Earth-CAPHAs both in terms of number and frequency, indicating tremendous prospects for future Mars-based observations.

It has been noted that a Mars-CAPHA may become an Earth-CAPHA later, and vice versa \citep{zhou2024}. Here, we find that it can also become a Venus-CAPHA or a Mercury-CAPHA. For example, in simulation~A1, a total of 41 Mercury-CAPHA, 44 Venus-CAPHA, 47 Earth-CAPHAs and 89 Mars-CAPHAs are recorded, with 41 of them being in common. This raises an intriguing possibility of using CAPHAs as matter carriers or transportation between terrestrial planets.

Moreover, the number of NEAs with $15\leq H\leq 22$ can also be predicted by our simulations. We define an asteroid to be in the NEA state when its perihelion distance $q\leq1.3$\,au. Similar to the (CAPHA) active lifetime for an asteroid in Section~\ref{sec:results_quantity}, its NEA active lifetime $t_{\rm act,NEA}$ is defined as the time difference between its first and last occurrence in the NEA state, and the NEA active coefficient is
\begin{equation}
    \epsilon_{\rm NEA}=\frac{t_{\rm act,NEA}}{T_{\rm sim}}.
\end{equation}
After correcting the number of NEAs in our simulations by coefficients $\epsilon_{\rm NEA}$ and $\eta$, we obtain a predicted value of 24766 for NEAs with $15\leq H\leq 22$. As of June 26, 2023, the observed number of such NEAs is 10468, indicating that many more are yet to be observed. By comparing the predicted numbers of CAPHAs and NEAs, it is found that 8\%, 12\%, and 15\% of NEAs can become CAPHAs of Mercury, Venus, and Earth, respectively. It should be noted that, due to Mars' perihelion distance $q=1.38$\,au, Mars-CAPHAs are not a subset of NEAs.

\textit{Previous work} \citep{zhou2024} uses a nominal timestep of 8 days, and it sets all asteroids in a `near-gap' region to drift toward the center of the gap (the increment caused by the Yarkovsky drift are halved when processing the data). While in \textit{this work}, a 4-hour nominal timestep is used, and the drift direction is randomized. To investigate the individual impact of these two settings on the results, we conduct an additional set of simulations with a 4-hour timestep and a drift direction toward the center (referred to as \textit{bridge simulation}). \textit{Previous work} predicts 4675 Earth-CAPHAs and 16910 Mars-CAPHAs, while \textit{bridge simulation} predicts 3767 Earth-CAPHAs and 15028 Mars-CAPHAs, both of which are fewer than those in \textit{previous work}. This is because the reduction in nominal timestep makes main belt asteroids behave more stably in the simulations, thereby reducing the occurrence of artifacts. \textit{This work} predicts 3791 Earth-CAPHAs and 18066 Mars-CAPHAs, which is more than \textit{bridge simulation}. This is because drifting in the direction opposite to the center of the belonging gap, toward a more distant gap, can also allow the asteroid to escape from the main belt, which is especially true for asteroids between 5:2 and 7:3 resonances (these two gaps are relatively close to each other).

\subsection{Relative velocities of CAPHAs}\label{sec:velocity}
We also investigate the distribution of relative velocities of asteroids during close encounters, and the results predicted by simulations are shown in Figure~\ref{fig:3} and Figure~\ref{fig:4}. The peak relative velocities of CAPHAs for Mercury, Venus, Earth, and Mars are about 24\,km\,s$^{-1}$, 18\,km\,s$^{-1}$, 12\,km\,s$^{-1}$, and 5\,km\,s$^{-1}$, respectively. This can be simply explained by the fact that when an asteroid gets closer to the Sun, more of its potential energy is converted into kinetic energy, resulting in a higher velocity. The simulation and observation of Earth-CAPHAs in Figure~\ref{fig:4} show a good agreement, indicating the reliability of our simulations. The lower relative velocities of Mars-CAPHAs suggest the good feasibility of Mars-based observations in the future.

It should be noted that when initially plotting the simulation velocity curve for Earth-CAPHAs in Figure~\ref{fig:4}, we also notice a small peak in the range of 25–35\,km\,s$^{-1}$. After carefully checking the simulation data, we find that this peak is `spurious' and is caused by nine anomalous asteroids, SPK-IDs (and corresponding simulations) of which are 20017248 (A4), 20067507 (A5), 20569467 (A5), 3851278 (A6), 3730267 (A7), 20382818 (A7), 20084635 (B6), 20502633 (B6), and 20108910 (E8). These nine asteroids all exhibit exceptionally long active lifetimes and experience thousands of high-velocity close encounters with Earth in the simulation. However, their perihelions reached $\sim$0.1\,au or even smaller at certain points (most close encounters occur after these moments), which would subject them to intense solar influence, potentially causing them to disintegrate or even melt (such mechanism is not included in our simulation). For instance, in simulation~E7, asteroid 20108910 had an orbital perihelion of only 0.02\,au at t=66.75\,Myr, after which it underwent 5411 high-velocity close encounters with Earth until t=1\,Gyr (the end of the simulation). Therefore, now the blue curve in Figure~\ref{fig:4} is the product after removing these nine `unrealistic' asteroids. Moreover, testing has shown that the numerous close encounters these nine asteroids had with Earth occurred by chance. If the Yarkovsky drift rate or some numerical parameters of the program were slightly altered, their trajectories would no longer exhibit this pattern. Additionally, since individual asteroids cannot significantly impact the total count, and our calculation of occurrence frequency is based on the average frequency per asteroid (see equation~(\ref{eq:frequency_f})), these nine asteroids have minimal effect on both the quantity and frequency.

\section{Summary}\label{sec:summary}
We incorporate an additional Yarkovsky effect subroutine into the N-body software package \textsc{Mercury6}, performing numerical simulations to study Mercury-CAPHAs and Venus-CAPHAs. Assuming that CAPHAs mainly originate from the main belt, 1\% of main belt asteroids are randomly sampled as the initial condition for our simulations. Given that `gaps' in the main belt are the primary escape routes for asteroids, we divide our sample into `near-gap' asteroids and `far-gap' asteroids, and run them with different configurations.

The numbers of Mercury-CAPHAs and Venus-CAPHAs are 1893 and 3014, respectively, with occurrence frequencies of 1 and 9 per year. The peak relative velocities of them are approximately 24\,km\,s$^{-1}$ and 18\,km\,s$^{-1}$, awaiting verification by future Mercury and Venus exploration missions. Our results for Earth-CAPHAs are in good agreement with observations, indicating the validity of our simulations. Additionally, the number of Mars-CAPHAs is abundant, and their relative velocities are relatively small, making them a promising target for future Mars-based exploration. The proportion of CAPHAs within NEAs is not insignificant, and studying them will enhance our understanding of terrestrial planet environments, asteroid-planet interactions, and the history of the Solar System. Moreover, a CAPHA of a planet can become that of another planet later, raising a possibility of using CAPHAs as transportation between different planets.

\section*{Acknowledgements}
This work is supported by the National Natural Science Foundation of China (NSFC, Grant Nos. 12225302, 12373081 and 12150009) and the CNSA program D050102.

\section*{Data Availability}
The data underlying this article will be shared on reasonable request to the corresponding author.
\bibliographystyle{mnras}
\bibliography{mercury} 
\appendix
\section{Table A1}
\begin{table*}
\footnotesize
	\centering
	\caption{The results for Earth-CAPHAs and Mars-CAPHAs. `EC' and `MaC' are abbreviations for Earth-CAPHA and Mars-CAPHA. Meanings of the elements here are similar to those in Table~\ref{tab:setting_result}.}
	\label{tab:setting_result_mars}
	\begin{tabular}{cccccccccc}
		\hline
		ID & drift rate (au\,Gyr$^{-1}$) & region information & $N_{\rm EC}$ & $N_{\rm MaC}$ & $\bar \epsilon_{\rm EC}$ & $\bar \epsilon_{\rm MaC}$ & $\eta$ & $N_{\rm EC,obs}$ & $N_{\rm MaC,obs}$ \\
            (1) & (2) & (3) & (4) & (5) & (6) & (7) & (8) & (9) & (10) \\
		\hline
        A1 & 0.18  & range (au): & 47 & 89 & 0.082 & 0.487 & 0.051 & 0.195 & 2.198\\
        A2 & 0.29 & $2.065^{+0.135}_{-0.050}$ & 59 & 100 & 0.064 & 0.446 & 0.115 & 0.432 & 5.129\\
        A3 & 0.46 & ~ & 48 & 96 & 0.099 & 0.457 & 0.261 & 1.234 & 11.450\\
        A4 & 0.72 & resonance: & 72 & 115 & 0.072 & 0.465 & 0.236 & 1.228 & 12.609\\
        A5 & 1.13 & 4:1 and $\nu_6$ & 61 & 112 & 0.113 & 0.480 & 0.355 & 2.448 & 19.091\\
        A6 & 1.82 & ~ & 78 & 133 & 0.056 & 0.349 & 0.535 & 2.319 & 24.809\\
        A7 & 2.83 & population: 198 & 81 & 126 & 0.074 & 0.366 & 0.806 & 4.854 & 37.160\\
        \hline
        B1 & 0.12 & range (au): & 44 & 44 & 0.034 & 0.041 & 0.079 & 0.117 & 0.143\\
        B2 & 0.20 & $2.502\pm 0.025$ & 61 & 61 & 0.026 & 0.057 & 0.198 & 0.315 & 0.694\\
        B3 & 0.31 & ~ & 75 & 76 & 0.015 & 0.031 & 0.497 & 0.571 & 1.181\\
        B4 & 0.49 & resonance: & 71 & 72 & 0.018 & 0.026 & 0.483 & 0.623 & 0.889\\
        B5 & 0.78 & 3:1 & 84 & 85 & 0.013 & 0.020 & 0.765 & 0.829 & 1.305\\
        B6 & 1.25 & ~ & 96 & 96 & 0.037 & 0.057 & 1.212 & 4.340 & 6.587\\
        B7 & 1.94 & population: 122 & 105 & 106 & 0.012 & 0.041 & 1.919 & 2.348 & 8.356\\
        \hline
        C1 & 0.10 & range (au): & 43 & 43 & 0.003 & 0.006 & 0.120 & 0.013 & 0.030\\
        C2 & 0.15 & $2.825\pm 0.020$ & 50 & 50 & 0.003 & 0.008 & 0.322 & 0.042 & 0.132\\
        C3 & 0.24 & ~ & 53 & 54 & 0.006 & 0.008 & 0.868 & 0.277 & 0.360\\
        C4 & 0.38 & resonance: & 68 & 68 & 0.006 & 0.012 & 0.885 & 0.389 & 0.696\\
        C5 & 0.60 & 5:2 & 73 & 73 & 0.003 & 0.006 & 1.453 & 0.329 & 0.592\\
        C6 & 0.96 & ~ & 101 & 101 & 0.008 & 0.014 & 2.384 & 1.935 & 3.320\\
        C7 & 1.50 & population: 132 & 106 & 106 & 0.004 & 0.006 & 3.913 & 1.469 & 2.603\\
        \hline
        D1 & 0.09 & range (au): & 14 & 31 & 0.000 & 0.004 & 0.120 & 0.001 & 0.016\\
        D2 & 0.14 & $2.958\pm 0.015$ & 20 & 33 & 0.002 & 0.006 & 0.322 & 0.012 & 0.062\\
        D3 & 0.22 & ~ & 20 & 36 & 0.004 & 0.015 & 0.868 & 0.074 & 0.454\\
        D4 & 0.35 & resonance: & 28 & 53 & 0.002 & 0.006 & 0.885 & 0.048 & 0.267\\
        D5 & 0.56 & 7:3 & 37 & 59 & 0.002 & 0.006 & 1.453 & 0.086 & 0.536\\
        D6 & 0.89 & ~ & 39 & 62 & 0.003 & 0.009 & 2.384 & 0.302 & 1.369\\
        D7 & 1.39 & population: 125 & 34 & 65 & 0.001 & 0.005 & 3.913 & 0.109 & 1.291\\
        \hline
        E1 & 0.07 & range (au): & 8 & 10 & 0.006 & 0.007 & 0.172 & 0.008 & 0.012\\
        E2 & 0.11 & $3.279\pm 0.050$ & 16 & 20 & 0.006 & 0.018 & 0.470 & 0.045 & 0.168\\
        E3 & 0.18 & ~ & 18 & 23 & 0.015 & 0.027 & 1.284 & 0.348 & 0.792\\
        E4 & 0.28 & resonance: & 24 & 27 & 0.007 & 0.018 & 1.322 & 0.237 & 0.628\\
        E5 & 0.44 & 2:1 & 19 & 29 & 0.010 & 0.032 & 2.185 & 0.431 & 2.009\\
        E6 & 0.71 & ~ & 27 & 30 & 0.008 & 0.019 & 3.611 & 0.826 & 2.079\\
        E7 & 1.11 & population: 77 & 24 & 34 & 0.063 & 0.063 & 5.968 & 9.067 & 12.807\\
		\hline
        G1 & 0 & zone \romannumeral1 & 0 & 54 & 0.000 & 0.344 & 0.337 & 0.000 & 6.250\\
        G2 & 0 & zone \romannumeral2 & 0 & 46 & 0.000 & 0.326 & 0.736 & 0.000 & 11.036\\
        G3 & 0 & zone \romannumeral3 & 0 & 6 & 0.000 & 0.182 & 1.421 & 0.000 & 1.548\\
        G4 & 0 & zone \romannumeral4 & 0 & 0 & 0.000 & 0.000 & 2.145 & 0.000 & 0.000\\
        G5 & 0 & zone \romannumeral5 & 1 & 1 & 0.003 & 0.000 & 3.288 & 0.011 & 0.000\\
		\hline
        total & - & - & - & - & - & - & - & 37.91 & 180.66\\
        \hline
	\end{tabular}
\end{table*}
\bsp	
\label{lastpage}
\end{document}